\documentclass{article}
\pdfoutput=1
\usepackage[onehalfspacing]{setspace}
\usepackage{fullpage}
\usepackage{sectsty} \sectionfont{\large}
\usepackage{amsmath,amsfonts}
\usepackage{graphicx,url,natbib} 

\DeclareMathOperator*{\var}{var}

\def\eatzero#1{\ifnum#1=0\relax\else{#1}\fi}
\def\eattwo#1#2{\relax}
\def\svndate$#1: #2-#3-#4 #5 #6 (#7) ${\def\fmtdate{#2--#3--#4}}
\svndate$Date: 2010-10-29 09:55:00 -0400 (Fri, 29 Oct 2010) $

\title{A Comparison of Methods for Computing Autocorrelation Time}
\author{Madeleine B. Thompson}
\date{\fmtdate}

\begin{document}

\begin{center}

{\small Technical Report No.\ 1007,
 Department of Statistics, University of Toronto}

\vspace*{0.6in}

{\Large\bf A Comparison of Methods for Computing Autocorrelation Time}\\[16pt]

{\large Madeleine B. Thompson}\\[3pt]
 \url{mthompson@utstat.toronto.edu} \\
 \url{http://www.utstat.toronto.edu/mthompson} \\
October 29, 2010
\end{center}

\vspace*{6pt}

\begin{abstract}
This paper describes four methods for estimating autocorrelation
time and evaluates these methods with a test set of
seven series.  Fitting an autoregressive process appears to be the
most accurate method of the four.  An R package is provided for
extending the comparison to more methods and test series.
\end{abstract}

\section{Introduction}

Autocorrelation time measures the convergence rate of the sample
mean of a function of a stationary, geometrically ergodic Markov
chain with bounded variance.  Let $\{X_i\}_{i=1}^n$ be $n$ values
of a scalar function of the states of such a Markov chain, where
$EX_i = \mu$ and $\var X_i=\sigma^2$.  Let $\bar X_n$ be the sample
mean of $(X_1,\ldots,X_n)$.  Then, the autocorrelation time is the
value of $\tau$ such that:
\begin{equation}\label{clt}
\sqrt{\frac{n}{\tau}} \cdot \frac{\bar X_n - \mu}{\sigma}
  \Longrightarrow N(0,1)
\end{equation}
Conceptually, the autocorrelation time is the number of Markov chain
transitions equivalent to a single independent draw from the
distribution of $\{X_i\}$.  It can be a useful summary of the
efficiency of a Markov chain Monte Carlo (MCMC) sampler.  Since
there is rarely a closed-form expression for the autocorrelation
time, many methods for estimating it from the data itself have been
developed.  This document compares four of them: computing batch
means, fitting a linear regression to the log spectrum, summing an
initial sequence of the sample autocorrelation function, and modeling
as an autoregressive process.

For a broader context, two classic overviews of uncertainty in MCMC
estimation are \citet{geyer92} and \citet[\S6.3]{neal93}.  The most
popular alternative to autocorrelation time, potential scale
reduction, is discussed by \citet{gelman92}.

\section{Methods for computing the autocorrelation time}
\label{methods}

\subsection{Batch means}
\label{batch-means}

Equation~\ref{clt} is approximately true for any subsequence,
so we can divide the $\{X_i\}$ into batches of size
$m$ and compute the sample mean of each batch.  As $m$ goes to
infinity, the batch means
have asymptotic variance $\sigma^2\cdot\tau/m$.  So, if $s^2$ is
the sample variance of $\{X_i\}$ and $s_m^2$ is the sample variance
of the batch means, we can estimate the autocorrelation time with:
\begin{equation}\label{hattau-nm}
\hat\tau_{n,m} = m\frac{s_m^2}{s^2}
\end{equation}
For this to be a consistent estimator of $\tau$, the batch size and
the number of batches must go to infinity.  To ensure this, I use
$n^{1/3}$ batches of size $n^{2/3}$.  \citet[\S5.9]{fishman78}
discusses batch means in great detail; \citet[p.~104]{neal93} and
\citet[\S3.2]{geyer92} discuss batch means in the context of MCMC.

\subsection{Fitting a linear regression to the log spectrum}

In the spectrum fit method \citep[\S2.4]{heidelberger81}, a linear
regression is fit to the lower frequencies of the log-spectrum of the
chain and used to compute an estimate of the spectrum at frequency
zero, which we will denote by $\hat I_0$.  Let $s^2$ be the sample
variance of the $\{X_i\}$.  We can estimate the autocorrelation time with:
\begin{equation}
\hat\tau = \frac{\hat I_0}{s^2}
\end{equation}
This is implemented by the \texttt{spectrum0} function in R's CODA
package \citep{plummer06}.  First and second order polynomial fits
are practically indistinguishable on the test series described in
section~\ref{benchmark}, so this paper only explicitly includes the
results for a first order fit.

\subsection{Initial sequence estimators}

One formula for the autocorrelation time is \citep[p.~91]{straatsma86}:
\begin{equation}\label{tau}
\tau = 1 + 2\sum_{k=1}^\infty \rho_k
\end{equation}
where $\rho_k$ is the autocorrelation function (ACF) of the series 
at lag $k$.  For small lags, $\rho_k$ can be estimated with:
\begin{equation}\label{rhohat}
\hat\rho_k = \frac{1}{ns^2} \sum_{i=1}^{n-k} (X_i-\bar X_n)(X_{i+k}-\bar X_n)
\end{equation}
But, this is not defined for lags greater than $n-1$, and the partial
sum $\hat\rho_1 + \cdots + \hat\rho_{n-1}$ does not have a variance
that goes to zero as $n$ goes to infinity, so substituting all values
from equation~\ref{rhohat} into equation~\ref{tau} is not an
acceptable way to estimate $\tau$.

\citet{geyer92} shows that, for reversible Markov chains, sums of pairs
of consecutive ACF values, $\rho_i+\rho_{i+1}$, are always
positive, so one can obtain a consistent estimator, the initial
positive sequence (IPS) estimator, by truncating the sum when the
sum of adjacent sample ACF values is negative.

Further, the sequence of sums of pairs is always decreasing, so one
can smooth the initial positive sequence when a sum of two adjacent
sample ACF values is larger than the sum of the previous pair; the
resulting estimator is the initial monotone sequence (IMS) estimator.
Finally, the sequence of sums of pairs is also convex.  Smoothing
the sum to account for this results in the initial convex sequence
(ICS) estimator.

I was unable to distinguish the behavior of the IPS, IMS, and ICS
estimators, so the comparisons of this paper only directly include
the ICS estimator.

\subsection{AR process}

Another way to estimate the autocorrelation time, based on CODA's
\texttt{spectrum0.ar} \citep{plummer06}, is to model the series
as an AR($p$) process, with $p$ chosen by AIC.  We suppose a
model of the form:
\begin{equation}\label{armodel}
X_t = \mu + \pi_1 X_{t-1} + \cdots + \pi_p X_{t-p} + a_t,
\qquad a_t \sim N(0,\sigma_a^2)
\end{equation}
Let $\hat\rho_{1:p}$ be the vector $(\hat\rho_1,\ldots,\hat\rho_p)$.
We can obtain an estimate the AR coefficients, $\hat\pi_{1:p}$, with
the Yule--Walker method \citep[pp.~136--138]{wei06}.  Combining
equations 7.1.5 and 12.2.8b of \citet[pp.~137, 274--275]{wei06}, 
we can estimate the autocorrelation time with:
\begin{equation}\label{ar-tauhat}
\hat\tau = \frac{1-\hat\rho_{1:p}^T \hat\pi_{1:p}}
  {\left(1-1_p^T \hat\pi_{1:p}\right)^2}
\end{equation}
Because the Yule--Walker method estimates the asymptotic variance
of $\hat\pi_{1:p}$, Monte Carlo simulation can be used to generate
a confidence interval for the autocorrelation time.  For more
information on this method, see \citet[\S5.10]{fishman78} and
\citet{thompson10b}.

\section{Seven test series}
\label{benchmark}

\begin{figure}
\begin{center}\includegraphics{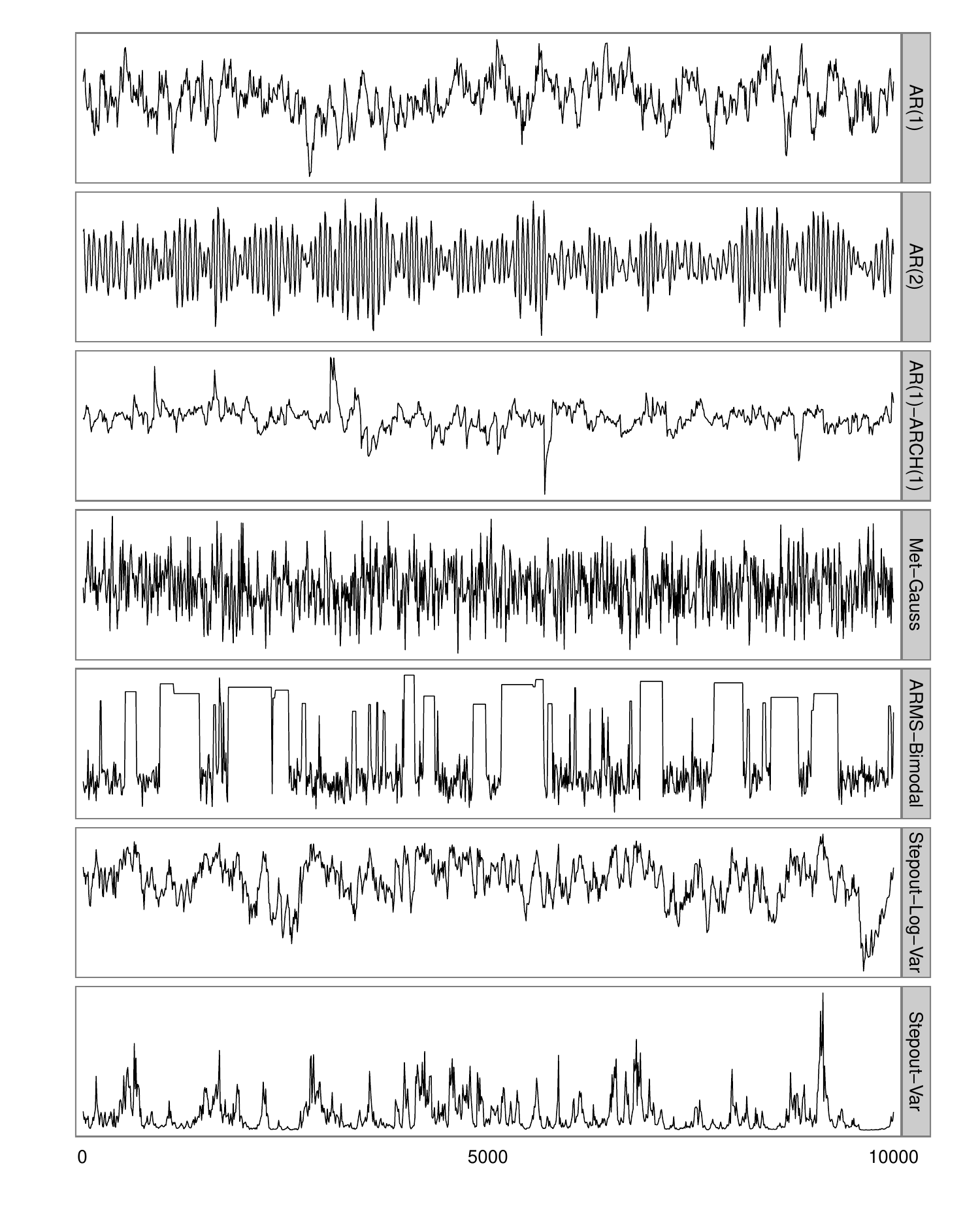}\end{center}
\caption{The first 10,000 elements of seven series used as
a test set.  See section~\ref{benchmark} for more information.}
\label{series}
\end{figure}

I evaluate the methods of section~\ref{methods} with seven series.
The first 10,000 elements of each are plotted in figure~\ref{series}.

\begin{itemize}
\item \textbf{AR(1)} is an AR(1) process with autocorrelation equal
to $0.98$ and uncorrelated errors.  Using equation~\ref{ar-tauhat},
its autocorrelation time is 99.  It is, in one sense, the simplest
possible series with an autocorrelation time other than one.

\item \textbf{AR(2)} is the AR(2) process with poles at $-1+0.1i$
and $-1-0.1i$:
\begin{equation}
Z_t = 1.98 \, Z_{t-1} - 0.99 \, Z_{t-2} + a_t, \quad a_t \sim N(0,1)
\end{equation}
Despite oscillating with a period of about 60, the terms in its
autocorrelation function nearly cancel each other out, so its
autocorrelation time is approximately two.  This series is included to
identify methods that cannot identify cancellation in long-range
dependence.

\item \textbf{AR(1)-ARCH(1)} is an AR(1) process with autocorrelation
$0.98$ and autocorrelated errors:
\begin{equation}
Z_t = 0.98 \, Z_{t-1} + a_t,
  \quad a_t \sim N\left(0, 0.01 + 0.99 \, a_{t-1}^2\right)
\end{equation}
Its autocorrelation time is also 99.  This series could
confound methods that assume constant step variance.

\item \textbf{Met-Gauss} is a sequence of states generated by a
Metropolis sampler with Gaussian proposals sampling from a Gaussian
probability distribution.  It has an autocorrelation time of eight.
It is a simple example of a state sequence from an MCMC sampler.

\item \textbf{ARMS-Bimodal} is also a sequence of states from an
MCMC simulation, ARMS \citep{gilks95} applied to a mixture of
two Gaussians.  The sampler is badly tuned, so it rarely accepts a
proposal when near the upper mode.  It has an autocorrelation time of approximately
200.

\item \textbf{Stepout-Log-Var} is a third MCMC sequence, the
log-variance parameter of slice sampling with stepping out
\citep[\S4]{neal03} sampling from a ten-parameter multilevel model
\citep[pp.~138--145]{gelman04}.  Its autocorrelation time is approximately 200.

\item \textbf{Stepout-Var} was generated by exponentiating the
states of Stepout-Log-Var.  Because its sample mean is dominated
by large observations, its autocorrelation time, approximately 100, is not the same
as that of Stepout-Log-Var.  Like ARMS-Bimodal and Stepout-Log-Var,
it is a challenging, real-world example of a sequence whose autocorrelation time
might be unknown.

\end{itemize}

\section{Comparison of methods}
\label{sec-comparison}

\begin{figure}
\begin{center}\includegraphics{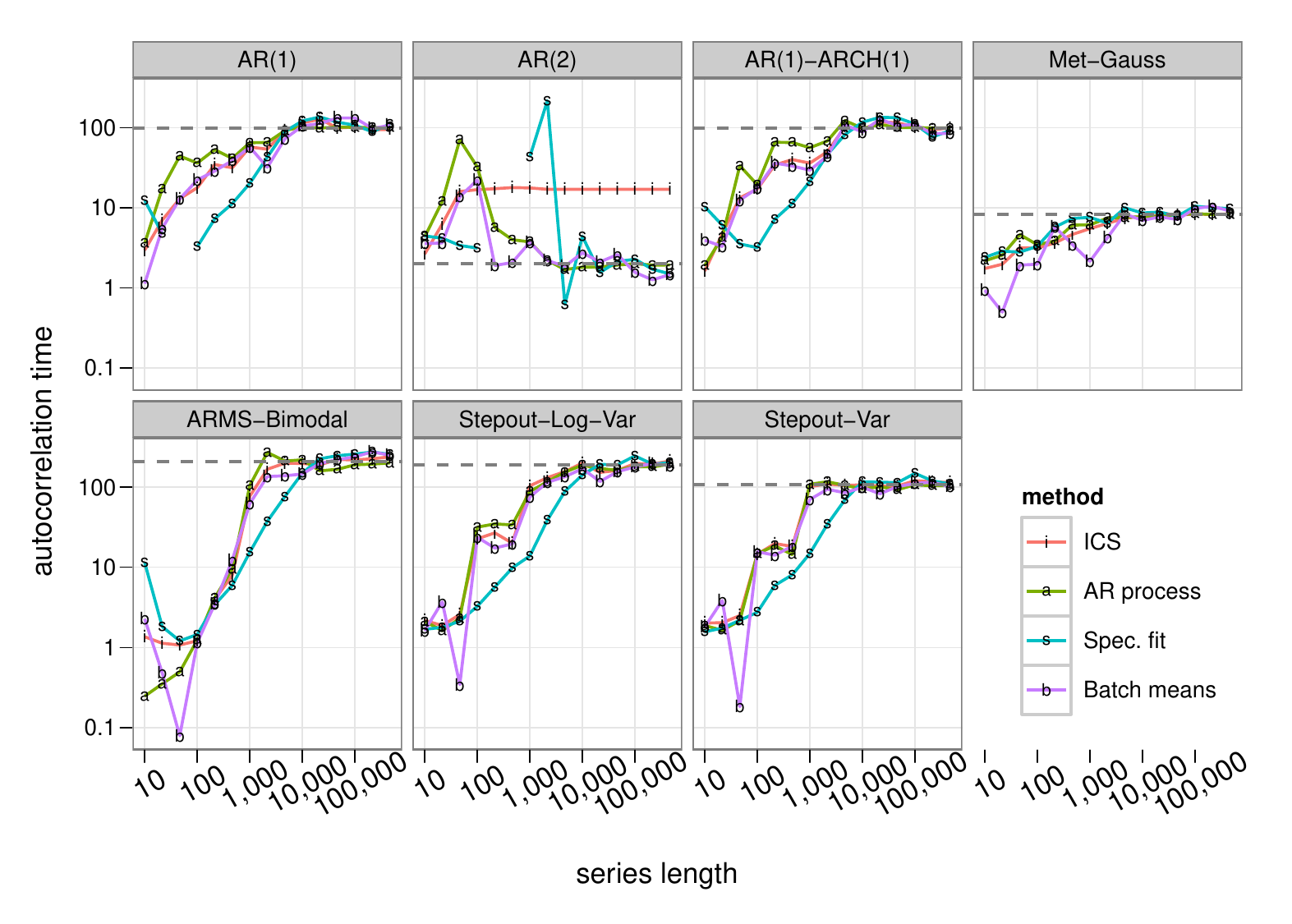}\end{center}
\caption{Autocorrelation times computed by four methods on seven
series for a range of subsequence lengths.  Each plot symbol
represents a single estimate.  The dashed line indicates the true
autocorrelation time.  Breaks in the lines for spectrum fit indicate
instances where the method failed to produce an estimate.  See
section~\ref{sec-comparison} for discussion.}
\label{fig-comparison}
\end{figure}

For each series, I compare the true autocorrelation time to the
autocorrelation time estimated by each method for subsequences
ranging in length from 10 to 500,000.  The results are plotted in
figure~\ref{fig-comparison}.  All four methods converge to the true
value on all seven chains, with the exception of ICS on the AR(2)
process.  In general, the AR process method converges to the true
autocorrelation time fastest, followed by ICS, batch means, and
finally spectrum fit, which does not even produce an estimate in
all cases.  All four methods tend to underestimate autocorrelation
times for all short sequences except those from the AR(2) process,
which appears in small samples to have a longer autocorrelation
time than it actually does.

\begin{figure}
\begin{center}\includegraphics{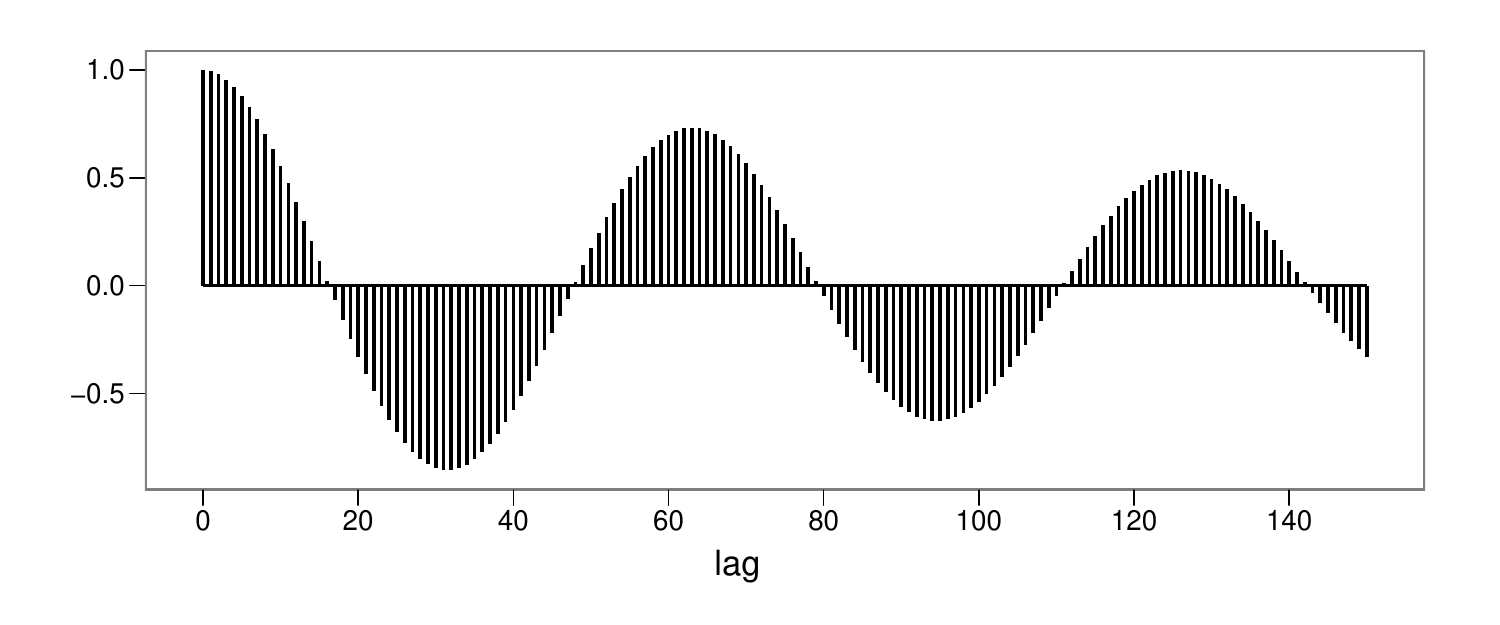}\end{center}
\caption{The autocorrelation function of the AR(2) series.  See
section~\ref{sec-comparison} for discussion.}
\label{ar2-acf}
\end{figure}

ICS is inconsistent on the AR(2) process because the autocorrelation
function of the AR(2) process is significant at lags larger than
its first zero-crossing (see figure~\ref{ar2-acf}).  ICS and the other
initial sequence estimators are not necessarily consistent when the
underlying chain is not reversible; an AR(2) process is not, when
considered a two state system.  These estimators stop summing the
sample autocorrelation function the first time its values fall below
zero, so these estimators never see that values at large lags cancel
values at small lags.

\begin{figure}
\begin{center}\includegraphics{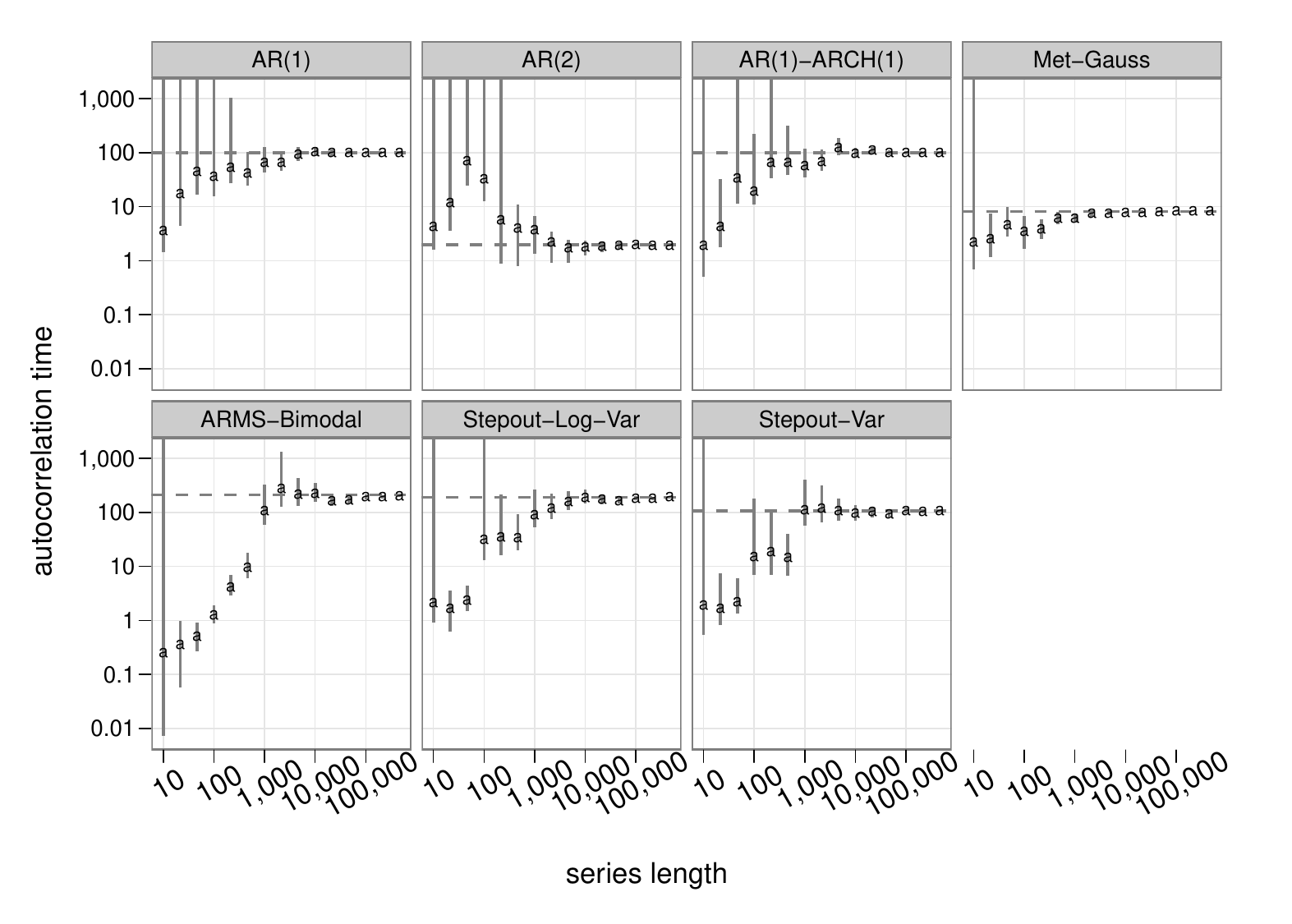}\end{center}
\caption{AR process-generated point estimates and 95\% confidence
intervals for the autocorrelation times of seven series for a range
of subsequence lengths.  The dashed line indicates the true
autocorrelation time.  The confidence intervals are somewhat
reasonable.  See section~\ref{sec-comparison} for discussion.}
\label{ar-ci}
\end{figure}

The AR process method can also generate approximate confidence
intervals for the autocorrelation time.  95\% confidence intervals
generated by the AR process method are shown in figure~\ref{ar-ci}.
The intervals are somewhat reasonable for subsequences longer than
the true autocorrelation time.  An exception is on AR(2), where the
first few hundred elements do not represent the entire series but
appear as though they might.

\section{Discussion}

The AR process method is more accurate than the three other methods
and is the only method that, as implemented, generates confidence
intervals, so I usually prefer it to the other three.  The batch
means estimate is faster to compute and almost as accurate, so it
may be useful when speed is important and confidence intervals are
not necessary.

The test set only contains seven series and reflects the type of
data I encounter in my own work.  To extend the comparison to other
methods and series, one can use ACTCompare, an R package available
at \url{http://www.utstat.toronto.edu/mthompson}.

\section{Acknowledgments}

I would like to thank Radford Neal for his comments on this paper
and Charles Geyer for his discussion of initial sequence estimators.

\bibliographystyle{apalike}
\bibliography{../corlen/corlen}

\begin{thebibliography}{}

\bibitem[Fishman, 1978]{fishman78}
Fishman, G.~S. (1978).
\newblock {\em Principles of Discrete Event Simulation}.
\newblock John Wiley and Sons.

\bibitem[Gelman et~al., 2004]{gelman04}
Gelman, A., Carlin, J.~B., Stern, H.~S., and Rubin, D.~B. (2004).
\newblock {\em Bayesian Data Analysis, Second Edition}.
\newblock Chapman and Hall/CRC.

\bibitem[Gelman and Rubin, 1992]{gelman92}
Gelman, A. and Rubin, D.~B. (1992).
\newblock Inference from iterative simulation using multiple sequences.
\newblock {\em Statistical Science}, (4):457--511.

\bibitem[Geyer, 1992]{geyer92}
Geyer, C.~J. (1992).
\newblock Practical {Markov} {Chain} {Monte} {Carlo}.
\newblock {\em Statistical Science}, 7(4):473--511.

\bibitem[Gilks et~al., 1995]{gilks95}
Gilks, W.~R., Best, N.~G., and Tan, K. K.~C. (1995).
\newblock Adaptive rejection {Metropolis} sampling within {Gibbs} sampling.
\newblock {\em Applied Statistics}, 44(4):455--472.

\bibitem[Heidelberger and Welch, 1981]{heidelberger81}
Heidelberger, P. and Welch, P.~D. (1981).
\newblock A spectral method for confidence interval generation and run length
  control in simulations.
\newblock {\em Communications of the ACM}, 24(4):233--245.

\bibitem[Neal, 1993]{neal93}
Neal, R.~M. (1993).
\newblock Probabilistic inference using {Markov Chain Monte Carlo} methods.
\newblock Technical Report CRG-TR-93-1, Dept. of Computer Science, University
  of Toronto.

\bibitem[Neal, 2003]{neal03}
Neal, R.~M. (2003).
\newblock Slice sampling.
\newblock {\em Annals of Statistics}, 31:705--767.

\bibitem[Plummer et~al., 2006]{plummer06}
Plummer, M., Best, N., Cowles, K., and Vines, K. (2006).
\newblock {CODA}: Convergence diagnosis and output analysis for {MCMC}.
\newblock {\em R News}, 6(1):7--11.

\bibitem[Straatsma et~al., 1986]{straatsma86}
Straatsma, T.~P., Berendsen, H. J.~C., and Stam, A.~J. (1986).
\newblock Estimation of statistical errors in molecular simulation
  calculations.
\newblock {\em Molecular Physics}, 57(1):89--95.

\bibitem[Thompson, 2010]{thompson10b}
Thompson, M.~B. (2010).
\newblock Graphical comparison of {MCMC} performance.
\newblock In preparation.

\bibitem[Wei, 2006]{wei06}
Wei, W. W.~S. (2006).
\newblock {\em Time Series Analysis: Univariate and Multivariate Methods,
  Second Edition}.
\newblock Pearson/Addison Wesley.

\end{thebibliography}

\end{document}